\documentclass{article}

\usepackage[dblblindworkshop, final]{neurips_2026}
\workshoptitle{Sim2Science: ML with Imperfect Scientific Models}

\usepackage[utf8]{inputenc} 
\usepackage[T1]{fontenc}    
\usepackage{url}            
\usepackage{booktabs}       
\usepackage{amsfonts}       
\usepackage{nicefrac}       
\usepackage{microtype}      
\usepackage{xcolor}         

\usepackage{booktabs}
\usepackage{tabularx}
\usepackage{array}
\usepackage{url}
\usepackage{enumitem}
\usepackage{graphicx}
\usepackage{amsmath, amssymb}
\usepackage{booktabs}
\usepackage[colorlinks=true, linkcolor=blue, urlcolor=blue]{hyperref}
\hypersetup{hidelinks}
\usepackage{fontawesome5}
\definecolor{yesgreen}{HTML}{238B45}
\definecolor{cautionamber}{HTML}{D99A00}
\newcommand{\yes}{\textcolor{yesgreen}{\faCheck}}
\newcommand{\caveat}{\textcolor{cautionamber}{\faExclamationTriangle}}

\newcommand{\fit}[1]{\textcolor{red}{#1}}

\newcommand{\Lc}{L_c}
\newcommand{\Nc}{N_c}
\newcommand{\Lr}{L_r}
\newcommand{\Nr}{N_r}
\newcommand{\xbarc}{\bar{x}_c}
\newcommand{\diff}{\mathrm{d}}

\title{Can LLM Agents Improve Imperfect Earth System Model Parameterizations?}

\author{%
  Kara D. Lamb\\
  Department of Earth and Environmental Engineering\\
  Learning the Earth with Artificial Intelligence and Physics Center\\
  Columbia University\\
  New York, NY 10027 \\
  \texttt{kl3231@columbia.edu} \\
}

\begin{document}

\maketitle

\begin{abstract}
Earth system models (ESMs) rely on parameterizations to represent unresolved processes such as convection, cloud microphysics, and turbulence. However, translating advances in physical understanding into stable, validated parameterizations and integrating them into ESMs often takes years. Agentic AI could accelerate this process and enable more systematic exploration of design choices. Using warm-rain collision–coalescence as a test case, we evaluate agents’ ability to synthesize scientific literature, reimplement published bulk parameterizations, and propose new functional forms as differentiable JAX code. We compare conventional parameterizations, data-driven emulators, and agent-generated closures against high-fidelity superdroplet simulations, repeating unattended experiments across ten language models. With expert guidance or iterative feedback, agents produced executable, differentiable, mass-conserving code, although unattended performance depended strongly on the scheme and language model. Iterative feedback lowered median rollout error for eight of ten models. Retuned physical schemes matched or outperformed many learned parameterizations at substantially lower computational cost, but two of 100 unattended agent-generated closures achieved slightly lower rain-mass error than the most accurate retuned scheme on held-out simulations.

\end{abstract}


\section{Introduction}
Earth System Models (ESMs) rely on parameterizations to represent unresolved processes such as clouds, turbulence, and ocean eddies. Recent advances in machine learning (ML) have enabled emulation of model components \cite[]{gentine2018could,rasp2018deep,bolton2019applications} and even full model surrogates \cite[]{allen2025end,bi2023accurate,lam2023learning,price2025probabilistic}. Despite these promising advances, physically grounded parameterizations remain essential for interpretability, scientific insight, and robustness for out-of-distribution climate prediction\cite[]{lamb_perspectives_2026}. However, it often takes years to translate new physical understanding into stable, validated parameterizations that can be integrated into ESMs. 

Although ML has been widely applied to individual modeling tasks, improving complex scientific models requires coordinated efforts involving literature review, data analysis, code development, calibration, and evaluation. Recent progress in AI agents and multi-agent AI systems offers a promising method to speed up this process \cite[]{tom_self-driving_2024,gottweis_towards_2025}. Specialized agents can synthesize scientific knowledge, analyze data, generate code, and evaluate candidate models in response to quantitative feedback \cite[]{openai_ai_2026}. Although agentic approaches are beginning to transform laboratory sciences, their ability to improve complex, high-dimensional computational models such as ESMs remains largely unexplored. Initial applications for ESM development have included the agentic translation of legacy Fortran model components into differentiable JAX code\cite[]{lahlou_systematic_2026} demonstrating these systems can assist with specific bottlenecks in model development. 

Here we explore whether agentic AI can be used to develop and systematically improve the sub-grid-scale parameterizations used in ESMs, going beyond translating existing code to explore where agentic AI can accelerate the model development process. As a proof of concept, we focus on warm-rain collision–coalescence, a process for which high-fidelity reference simulations and a range of established physical and machine-learning parameterizations provide a controlled test bed. We evaluate whether agentic workflows can synthesize information from the literature, translate existing schemes into differentiable JAX code, propose and refine new closures, calibrate model parameters, and compare competing approaches within a common framework. By comparing expert-guided, single-shot, and iterative workflows, we differentiate the ability to generate functional code from the more demanding task of producing physically consistent, numerically stable, and more accurate (with respect to the reference simulations) parameterizations. This analysis highlights where agentic AI can already accelerate  parameterization development, and where expert intervention remains essential. 

We ask three questions: (1) Can LLM agents reliably translate published parameterizations into accurate, differentiable code? (2) Does automated evaluation and feedback improve agent-generated closures? and (3) Do agent-generated or published learned closures outperform retuned physical parameterizations?

\begin{figure}
\noindent\includegraphics[width=\textwidth]{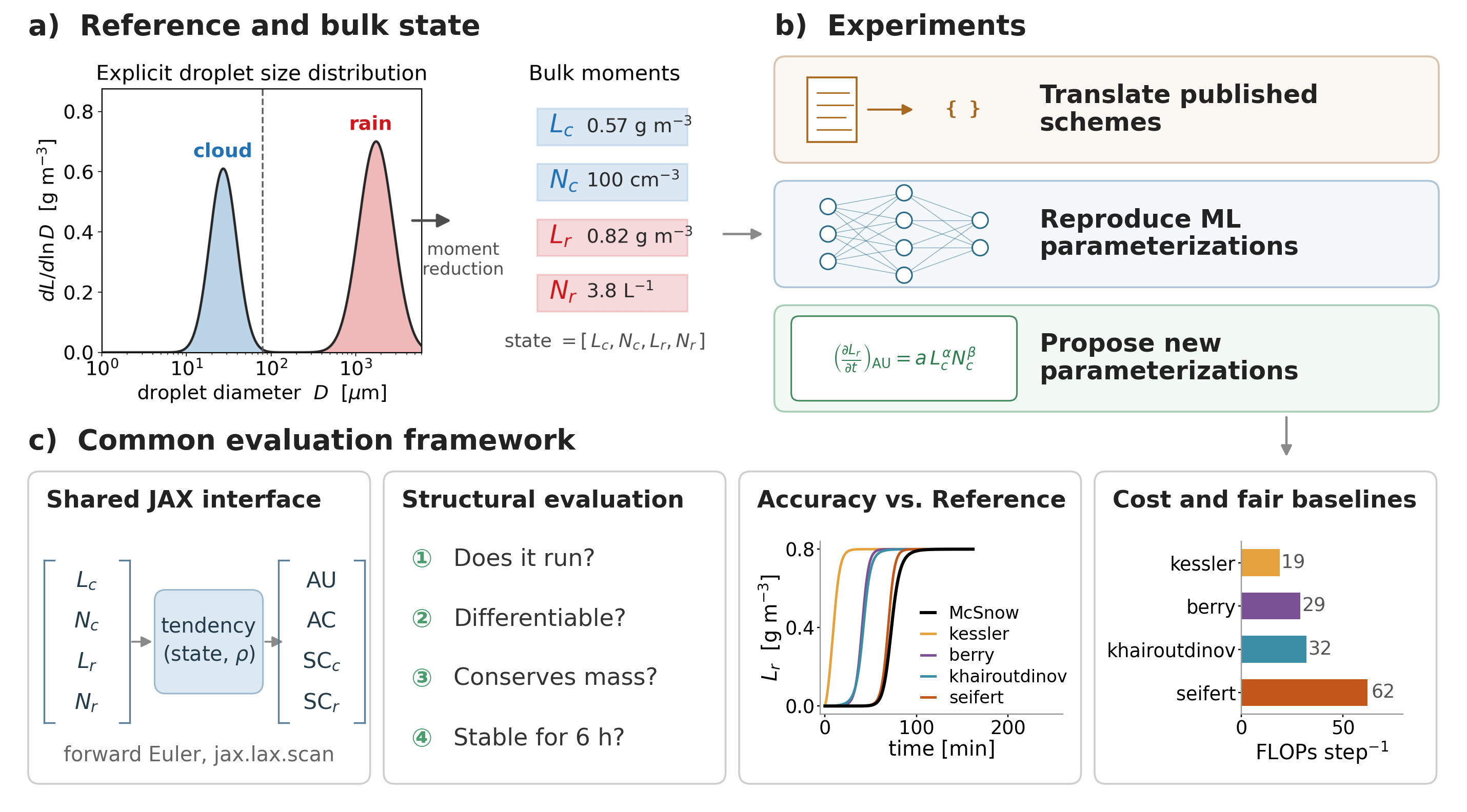}
\caption{\textbf{Overview of the analysis and methodology.}}
\label{fig:overview}
\end{figure}

\vspace{-0.5cm}
\section{An Atmospheric Cloud Parameterization Test Case}
\vspace{-0.3cm}
As a test case, we focus on warm-rain collision--coalescence, an important cloud microphysical process controlling precipitation formation, which is typically represented in ESMs using simplified bulk parameterizations (Figure \ref{fig:overview}a). In a two-moment scheme, the cloud (non-precipitating) and rain (precipitating) droplet populations are each represented by their grid-averaged number and mass mixing ratios, giving four prognostic variables $(N_c,L_c,N_r,L_r)$. Their evolution is described by four parameterized process rates (autoconversion, accretion, cloud self-collection, and rain self-collection), expressed as four coupled ODEs that conserve total mass. Classical schemes use analytic or empirically derived expressions for these rates \citep[e.g.][]{kessler1969distribution,berry1974analysis,khairoutdinovNewCloudPhysics2000,seifert2001double}, while recent approaches have used neural network regression \citep[]{seifert2020potential,sharma_superdropnet_2024}.

Because time-resolved observations of collision--coalescence inside clouds are not currently technologically possible \citep[]{morrison_confronting_2020}, we evaluate the parameterizations against high-fidelity McSnow simulations\cite[]{brdar2018mcsnow}. McSnow explicitly represents the stochastic collision-coalescence process using Lagrangian superdroplets, providing a high-fidelity reference that is too computationally expensive for use in ESMs. We use the dataset developed by \citet{sharma_superdropnet_2024}, which contains 819 initial conditions: 719 for training and validation and 100 for testing. The McSnow configuration, initial-conditions, and additional details are provided in the Appendix.

\vspace{-0.5cm}
\section{Evaluation of Agentic AI for Parameterization Development}
\vspace{-0.3cm}
To explore agentic translation of historical schemes into JAX, reproduction of existing ML closures, and the proposal of new closures (Figure \ref{fig:overview}b), we evaluate three different paradigms for LLM code development: human-in-the-loop agentic AI, single-shot API calls to an LLM, and iterative API calls with feedback. In the first, an expert remains actively involved in guiding the agent and directing iterative refinements. In the latter two, we use controlled single-shot and iterative API calls to evaluate what an LLM can accomplish without human intervention. Expert-guided experiments used Claude Opus~4.8, and unattended experiments were run with ten different language models (Appendix~\ref{appendix:models}), so differences between the two settings reflect both expert supervision and variations across LLM models. The resulting implementations are evaluated for agreement with reference schemes and compared in a common evaluation framework (Figure \ref{fig:overview}c).
\vspace{-0.3cm}
\paragraph{Expert-in-loop literature scheme translation}
We first evaluate whether agentic AI can translate established collision--coalescence parameterizations into a common JAX framework. We selected 4 traditional parameterizations that span a range of formulations and levels of complexity: \cite[]{kessler1969distribution,berry1974analysis,khairoutdinovNewCloudPhysics2000,seifert2001double}. Because these schemes use different prognostic variables and include both single- and two-moment representations of the droplet size distributions, their implementation requires more than a direct translation of individual equations. The agent must reconcile variable definitions, units, and process-rate conventions while mapping each scheme onto a standardized model interface and state representation. Additional details of these parameterizations and their functional forms are given in the Appendix.

Using expert-in-the-loop development, we constructed a JAX-native bulk microphysics library containing differentiable implementations of all four schemes. Very little expert guidance was required, with only 5 total prompts and $<$ 1 hour of active time required to translate all 4 schemes (Table \ref{tab:interventions}). All 4 resulting parameterizations were differentiable, mass conserving, and stable during prognostic integration. These expert-supervised implementations are used as reference baselines for evaluating the accuracy and reliability of the less closely supervised LLM translations.
\vspace{-0.3cm}
\paragraph{Expert-in-the-loop reconstruction of machine-learning baselines}
We next used expert-in-the-loop agentic development to reconstruct two machine learned collision--coalescence parameterizations in JAX: the process-rate neural networks of \citet{seifert2020potential} and SuperdropNet \citep[]{sharma_superdropnet_2024}. Given the original publications and McSnow training data, the agent closely reproduced the single-step and prognostic behavior of the Seifert and Rasp model. For SuperdropNet, it reconstructed the architecture and loss function but did not reproduce the reported autoregressive performance. The primary difficulty was reproducing the optimization procedure and training curriculum rather than implementing the network itself. The two reconstructions required 4.0 and 7.1 hours of active expert involvement, respectively (Table \ref{tab:interventions}). Additional methodological details and reconstructed evaluation figures are provided in Appendix~\ref{app:ml_reconstruction}.

\vspace{-0.3cm}
\paragraph{Expert-in-the-loop development of a novel physics-constrained Neural ODE closure}
We also used an expert-in-the-loop workflow to develop a physics-constrained Neural ODE for the four bulk moments \citep[]{chenNeuralOrdinaryDifferential2019}. The agent proposed an architecture that represented the learned tendencies through physically interpretable process rates and enforced mass conservation by construction. Development required substantial expert involvement: 131 prompts and approximately 22 active hours (Table \ref{tab:interventions}). Although the resulting model captured the qualitative evolution of the McSnow trajectories, it remained less accurate than several classic and machine learned baselines (Section \ref{sec:evaluation}). The agent contributed useful architectural ideas by recombining principles from existing physical schemes, but did not produce a superior parameterization.

\vspace{-0.3cm}
\paragraph{Single-shot API calls for historic parameterization reconstruction}
We tested whether the four literature parameterizations could be implemented through unattended, single-shot LLM calls. For each scheme, each of ten language models (Appendix~\ref{appendix:models}) received a standardized prompt, the original publication as a PDF, and the target JAX interface. We performed ten independent calls per scheme and model with no shared context, allowing us to evaluate variability in translating scientific descriptions into executable code.

Each implementation was tested for successful execution, automatic differentiation, mass conservation, and stability over a six-hour box-model rollout (Figure \ref{fig:overview}c). Successful implementations were then compared with the expert-supervised JAX library and McSnow reference simulations. Performance depended strongly on the scheme and model (Figure \ref{fig:sshistorical}), with the family of Claude models performing best at the translation task. Across the ten models, Kessler and Khairoutdinov \& Kogan translations passed all structural tests in 99/100 and 94/100 cases, compared with 85/100 for Seifert \& Beheng and 74/100 for Berry \& Reinhardt. For Berry \& Reinhardt, the models received Parts~I and II of the series \citep{berry1974analysis}, which describe the numerical method and the evolution of the drop spectrum; the rate equations appear in Part~IV \citep{berry1974analysis4}, which was not provided. Its lower pass rate therefore partly reflects missing information in the inputs, and its accuracy is measured against our Long-kernel implementation (Appendix) rather than the Part~IV equations.

\vspace{-0.3cm}
\paragraph{Automated LLM closure generation with iterative feedback.} 

We next tested whether LLMs could develop parameterizations autonomously using limited examples and automated feedback. Each model received a small sample of McSnow results, a common set of input variables, and physical and computational requirements. We compared single-shot generation with iterative refinement, in which models revised their candidates using feedback from training simulations. Final performance was assessed on 100 held-out simulations that were never used for feedback or candidate selection.

Across ten language models, iterative refinement over four rounds reduced median rain-mass error for eight models, although the improvement was statistically significant for only one (Figure~\ref{fig:iterativecomparison}). For half of the models, the best of four independent attempts performed at least as well as the median iteratively refined candidate. Differences among language models were larger than differences between single-shot and iterative performance. Nevertheless, some models consistently produced accurate closures: overall, 19 of 100 iteratively refined candidates outperformed the most accurate published physical scheme, Seifert \& Beheng, including seven of ten candidates from Claude Opus~5.5, and two, both from Claude Opus 5.5, outperformed Seifert \& Beheng retuned against McSnow (Appendix~\ref{appendix:models}). The best performing Claude scheme retained the Seifert \& Beheng autoconversion moment dependence but modified the correction functions and number budgets (Table \ref{tab:opus_vs_sb}).

Extending refinement from four to twelve rounds improved performance for both Claude Opus 5.5 and Gemini 4 Argon (Figure \ref{fig:iterativecomparison12rounds}). The larger gain occurred for Gemini, whose shorter searches frequently failed to produce a working closure, and the best candidate from each model outperformed published Seifert \& Beheng, with Claude's best candidate out-performing the retuned scheme. These results suggest that automated feedback can help LLMs discover accurate parameterizations, but its effectiveness depends strongly on the language model and the duration of the search.

\vspace{-0.5cm}
\section{Systematic Comparison of Implemented Parameterizations}\label{sec:evaluation}
\vspace{-0.3cm}
Implementing all parameterizations within a common JAX framework allows them to be compared systematically using the same state representation, reference simulations, numerical integration, and metrics. We evaluate both rain-formation timing and the complete rollout. The metric $t_{10}$ measures when 10\% of the initial cloud mass has been converted to the rain category (in minutes), while rollout accuracy is measured using mean absolute percentage error (MAPE) across $(N_c,L_c,N_r,L_r)$ on the held-out McSnow test set; MAPE for $L_r$ is shown in Figure \ref{fig:performance}. Further details are in Appendix~\ref{section:metrics}.

Because classical schemes were calibrated for particular reference models and initial conditions, their published coefficients may not provide the best achievable performance on the McSnow data set. We therefore retuned each scheme on the training set and evaluated both published and retuned versions on 100 test simulations, to separate the limitations of the functional form from those of its original calibration. Retuning substantially improved most classical schemes. We compared them with the expert-guided JAX implementations of Seifert and Rasp, SuperdropNet, and the physics-constrained NODE, as well as the two best-performing invented schemes (from Claude Opus 5.5). Advantages of the machine learned models over the published schemes often narrowed or disappeared after retuning, demonstrating the importance of comparing learned closures with recalibrated physical baselines. 

\begin{figure}
\noindent\includegraphics[width=\textwidth]{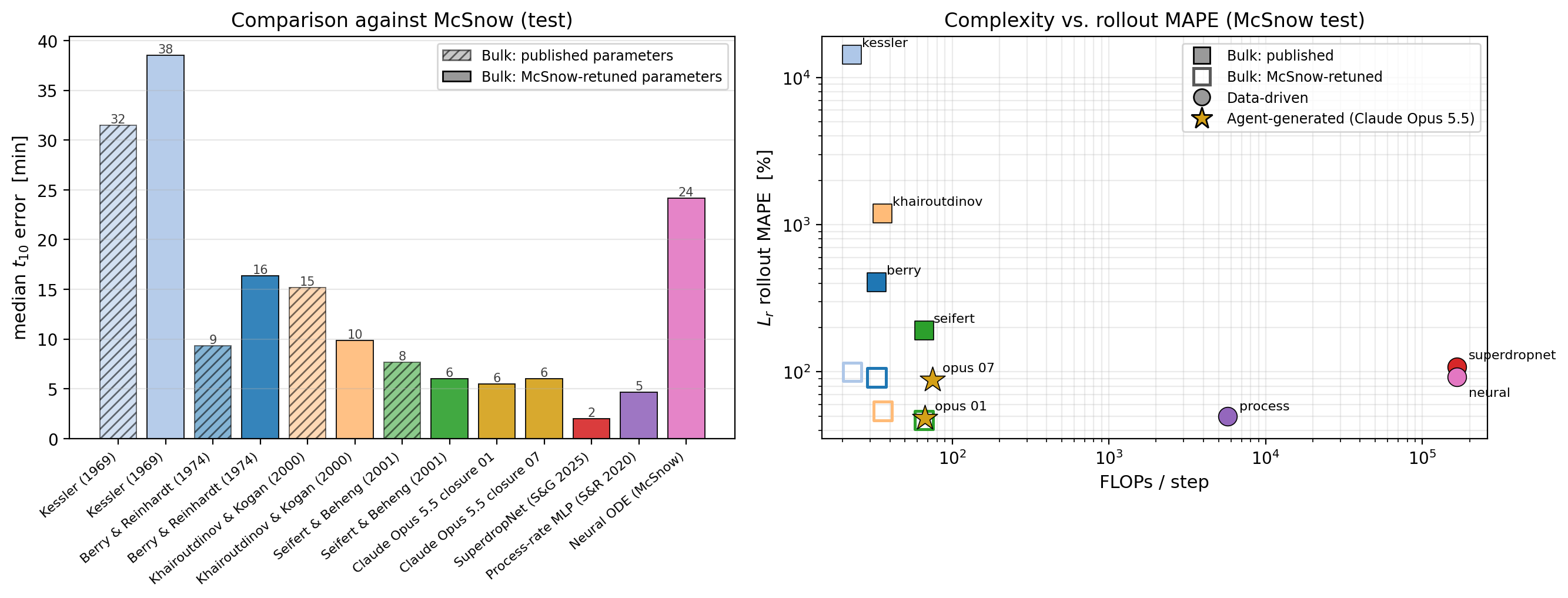}
\caption{\textbf{Parameterization performance.} (a) Mean error in rain-conversion timescales. Published and retuned coefficients are shown for the four literature schemes. (b) Rain-mass rollout error versus computational complexity, measured in floating-point operations per model step. Here, Superdropnet and Process-rate MLP refer to our reimplemented versions.}
\label{fig:performance}
\end{figure}

The suitability of a parameterization for integration in an ESM depends not just on accuracy, but also on numerical stability, physical consistency, computational cost, and robustness outside the training distribution. We quantify complexity using floating-point operations per model step, a hardware-independent measure that does not capture memory use or realized runtime. Several retuned physical schemes achieved errors comparable to the learned closures at substantially lower cost, although SuperdropNet most accurately captured the timing of the cloud-to-rain transition (Figure \ref{fig:performance}). No single scheme is uniformly superior across all evaluation criteria. This comparison illustrates the value of treating parameterization development as a systematic model-selection problem. 

\vspace{-0.5cm}
\section{Discussion}
\vspace{-0.3cm}

Physically grounded parameterizations remain essential for providing the interpretability, physical insight, and robustness required for climate prediction. The results presented here demonstrate that agentic AI can reconstruct historical parameterizations as differentiable JAX code, reconstruct machine learned baselines, and propose new physics-constrained closures. However, the agents were generally more successful at satisfying structural requirements such as executability, differentiability, mass conservation, and numerical stability than at improving predictive accuracy (Table \ref{table:answers}). These capabilities could accelerate the currently slow cycle of parameterization development, calibration, and evaluation, while making comparisons among alternative formulations more systematic and reproducible. The next step is to extend this evaluation beyond an isolated collision--coalescence box model. Recently developed JAX-based models such as the JAX Circulation Model or LegoESM provide a pathway for testing compatible implementations across a hierarchy ranging from process-level benchmarks and single-column models to intermediate complexity climate models and, eventually, coupled Earth system simulations \cite[]{davenport_jcm_2026,gentine2026legoesm}. Performance at the process scale does not guarantee good performance after coupling, where interactions with dynamics, thermodynamics, and other processes may amplify small errors or change the relative ranking of competing schemes. Although we focus here on one process, the workflow is not specific to it: any parameterization that can be written as a differentiable mapping from the resolved state to tendencies and evaluated against reference simulations can be developed in the same way. 

Closing the loop between parameterization generation, integration, evaluation, and revision could provide a foundation for increasingly self-improving models. Such models could systematically incorporate new observations, identify conditions for which available reference data are inadequate, design targeted high-fidelity simulations, compare alternative physical and data-driven formulations, and use the resulting evidence to revise their parameterizations. The present study addresses only a subset of this cycle, but demonstrates several of the capabilities needed to support it. However, important challenges still remain. It is not yet clear whether agentic AI can generate genuinely improved physical representations rather than primarily recombining ideas from existing approaches. Agentic development can also produce increasingly large and complex code bases containing subtle or hallucinated errors that may be difficult to identify. Agent-generated parameterizations must therefore meet the same requirements for physical consistency, numerical stability, reproducibility, and scientific validation as human-generated code. For parameterizations incorporated into ESMs used for weather forecasting or climate adaptation and mitigation decisions, additional safeguards should include independent review, automated conservation and stability tests, and out-of-distribution evaluation. 

\begin{ack}
We acknowledge funding and Google Cloud Platform support from a Google Award for Machine Learning Research \& Education with TPUs. We also acknowledge funding and computational resources from the U.S. Department of Energy, Office of Science, Office of Biological and Environmental Research, as part of the Genesis Mission National Science and Technology Challenges under Award \# DE-SC0026566. We acknowledge support from NSF through the Learning the Earth with Artificial Intelligence and Physics (LEAP) Science and Technology Center (STC) (Award \#2019625). 
\end{ack}

\newpage 

\section*{NeurIPS Paper Checklist}

\begin{enumerate}

\item {\bf Claims}
    \item[] Question: Do the main claims made in the abstract and introduction accurately reflect the paper's contributions and scope?
    \item[] Answer: \answerYes{}{} 
    \item[] Justification: The main claims are supported by our analysis.

\item {\bf Limitations}
    \item[] Question: Does the paper discuss the limitations of the work performed by the authors?
    \item[] Answer: \answerYes{} 
    \item[] Justification: We discuss both the performance and failures of our current approaches.

\item {\bf Theory assumptions and proofs}
    \item[] Question: For each theoretical result, does the paper provide the full set of assumptions and a complete (and correct) proof?
    \item[] Answer: \answerNA{} 
    \item[] Justification: This paper does not include theoretical results.

    \item {\bf Experimental result reproducibility}
    \item[] Question: Does the paper fully disclose all the information needed to reproduce the main experimental results of the paper to the extent that it affects the main claims and/or conclusions of the paper (regardless of whether the code and data are provided or not)?
    \item[] Answer: \answerYes{} 
    \item[] Justification: We have provided details of the LLMs used, the experiments performed, and the relative performance of different approaches.

\item {\bf Open access to data and code}
    \item[] Question: Does the paper provide open access to the data and code, with sufficient instructions to faithfully reproduce the main experimental results, as described in supplemental material?
    \item[] Answer: \answerNo{} 
    \item[] Justification: Since this is a workshop submission, we plan to release code upon acceptance of an expanded version of this paper in a peer-reviewed subject matter journal.

\item {\bf Experimental setting/details}
    \item[] Question: Does the paper specify all the training and test details (e.g., data splits, hyperparameters, how they were chosen, type of optimizer) necessary to understand the results?
    \item[] Answer: \answerYes{}{} 
    \item[] Justification: Details of the training and test data is given in the main paper and appendix.

\item {\bf Experiment statistical significance}
    \item[] Question: Does the paper report error bars suitably and correctly defined or other appropriate information about the statistical significance of the experiments?
    \item[] Answer: \answerYes{} 
    \item[] Justification: Information about error bars are provided in the relevant figure captions.

\item {\bf Experiments compute resources}
    \item[] Question: For each experiment, does the paper provide sufficient information on the computer resources (type of compute workers, memory, time of execution) needed to reproduce the experiments?
    \item[] Answer: \answerYes{} 
    \item[] Justification: Time and computational resources are explicitly evaluated in the paper.
    
\item {\bf Code of ethics}
    \item[] Question: Does the research conducted in the paper conform, in every respect, with the NeurIPS Code of Ethics \url{https://neurips.cc/public/EthicsGuidelines}?
    \item[] Answer: \answerYes{} 
    \item[] Justification: The research conducted in this paper conforms to the NeurIPS Code of Ethics.

\item {\bf Broader impacts}
    \item[] Question: Does the paper discuss both potential positive societal impacts and negative societal impacts of the work performed?
    \item[] Answer: \answerYes{} 
    \item[] Justification: Caveats around the need to rigorously evaluate code for ESMs is discussed in the Discussion.
    
\item {\bf Safeguards}
    \item[] Question: Does the paper describe safeguards that have been put in place for responsible release of data or models that have a high risk for misuse (e.g., pre-trained language models, image generators, or scraped datasets)?
    \item[] Answer: \answerNA{} 
    \item[] Justification: The paper poses no such risks.

\item {\bf Licenses for existing assets}
    \item[] Question: Are the creators or original owners of assets (e.g., code, data, models), used in the paper, properly credited and are the license and terms of use explicitly mentioned and properly respected?
    \item[] Answer: \answerYes{} 
    \item[] Justification: Yes, data and models are properly credited.

\item {\bf New assets}
    \item[] Question: Are new assets introduced in the paper well documented and is the documentation provided alongside the assets?
    \item[] Answer: \answerYes{} 
    \item[] Justification: Model code will be released upon publication of these results in a peer-reviewed journal. Data is accessible via the original SuperdropNet paper.

\item {\bf Crowdsourcing and research with human subjects}
    \item[] Question: For crowdsourcing experiments and research with human subjects, does the paper include the full text of instructions given to participants and screenshots, if applicable, as well as details about compensation (if any)? 
    \item[] Answer: \answerNA{} 
    \item[] Justification: Neither crowdsourcing nor human subjects were involved.

\item {\bf Institutional review board (IRB) approvals or equivalent for research with human subjects}
    \item[] Question: Does the paper describe potential risks incurred by study participants, whether such risks were disclosed to the subjects, and whether Institutional Review Board (IRB) approvals (or an equivalent approval/review based on the requirements of your country or institution) were obtained?
    \item[] Answer: \answerNA{} 
    \item[] Justification: Neither crowdsourcing nor human subjects were involved.

\item {\bf Declaration of LLM usage}
    \item[] Question: Does the paper describe the usage of LLMs if it is an important, original, or non-standard component of the core methods in this research? Note that if the LLM is used only for writing, editing, or formatting purposes and does \emph{not} impact the core methodology, scientific rigor, or originality of the research, declaration is not required.
    \item[] Answer: \answerYes{}
    \item[] Justification: Claude Code Opus 4.8 was used in the generation of code (as described in the manuscript) with human-in-the-loop oversight. Single-shot and iterative API calls to ten models were used to evaluate typical LLMs failure modes. Claude Code and ChatGPT were given access to the draft manuscript to provide targeted criticism and feedback based on the implemented code base. 

\end{enumerate}
\newpage
\appendix

\section{Results of experiments}
\begin{table}[h]\label{table:answers}
\centering\small
\begin{tabularx}{\textwidth}{@{}p{5.2cm} c X@{}}
\toprule
Research question & & Answer \\
\midrule
(1) Can LLM agents translate published schemes into differentiable code? & \yes &
Yes with light expert guidance; unattended success depends on the scheme and the
model. \\
\addlinespace
(2) Does automated feedback improve agent-generated closures? & \caveat &
Structural validity improves reliably; accuracy gains are modest and depend on the
model. Twelve-round searches improved performance for the two models examined. \\
\addlinespace
(3) Do agent-generated or learned closures outperform retuned physical schemes? &
\caveat & Two agent-generated closures do in rain-mass error, both from Claude Opus 5.5; learned closures only for the timing of rain onset. \\
\bottomrule
\end{tabularx}
\caption{Answers to the three research questions. \yes~yes; \caveat~yes, with
caveats.}
\end{table}

\section{High fidelity reference simulations}\label{appendix:McSnow}

As described in the main text, box model simulations using the McSnow Superdroplet Microphysics scheme \cite[]{brdar2018mcsnow} are used as high fidelity reference schemes for the collision-coalescence process. These data sets are stored as arrays with dimensions
\begin{equation}
    (\mathrm{time},\mathrm{feature},\mathrm{initial\ condition},
    \mathrm{ensemble\ member})
    =
    (3599,18,N,100),
\end{equation}
where $N=719$ for the training and validation data and $N=100$ for the held-out test data. Model output is recorded every $\Delta t=20$~s, and each initial condition contains 100 stochastic ensemble members; here we average over the 100 ensemble members for each set of initial conditions before fitting or optimizing parameters against the McSnow simulations. The 18 features include six cloud and rain moments, $(L_c,N_c,Z_c,L_r,N_r,Z_r)$; diagnosed number- and mass-transfer rates associated with autoconversion and accretion; self-collection diagnostics; and the initial liquid water content, mean cloud-droplet radius, and droplet-distribution shape parameter, $(\mathrm{lwc}_0,r_{c,0},\nu_0)$. Because the time required to convert cloud droplets to rain varies among initial conditions, shorter trajectories are padded to the maximum array length. A corresponding mask is applied so that padded values are excluded from training and evaluation. 

The reference simulations use a hydrodynamic gravitational collection kernel,
\begin{equation}\label{eq:hydro_kernel}
    K(x,y)=\pi\left[r(x)+r(y)\right]^2\left|v(x)-v(y)\right|E_{\mathrm{coll}}(x,y),
\end{equation}
where $x$ and $y$ are droplet masses, $r(x)$ is the corresponding droplet radius, $v(x)$ is the terminal fall velocity, and $E_{\mathrm{coll}}(x,y)$ is the tabulated collision efficiency of \citet{hall1980detailed} with the modifications of \citet{beheng1994parameterization}. The geometric cross section and terminal-velocity difference describe the volume swept out by droplets of different sizes, while $E_{\mathrm{coll}}$ accounts for the fraction of hydrodynamic encounters that result in a collision. 

Each simulation is initialized by sampling droplets of total mass $L_0$ from a generalized gamma distribution with shape parameter $\nu$ and mean droplet radius $\bar{r}_0$; $\nu = 0$ recovers an exponential distribution. Three
parameters are varied, on a full factorial grid (Table~\ref{tab:ics}). The box volume ($2500\,\mathrm{m^3}$) and the superdroplet multiplicity ($26500$) are held fixed, as is the gamma DSD family.

\begin{table}[h]
  \centering
  \caption{Initial conditions varied in the McSnow box-model ensemble. 
  The grid is $13 \times 7 \times 9 = 819$ distributions, split into 
  719 training and 100 test cases, each run as a 100-member stochastic 
  ensemble. Values follow Sharma \& Greenberg (2025, Table~1), whose 
  initial-condition set is inherited unchanged from Seifert \& Rasp (2020).}
  \label{tab:ics}
  \small
  \begin{tabular}{p{3.5cm}p{1.3cm}p{7.8cm}}
    \toprule
    Quantity & Symbol & Values \\
    \midrule
    Initial liquid water content
      & $L_0$ 
      & $\{0.2, 0.3, 0.4, 0.5, 0.6, 0.7, 0.8, 0.9,
        1.0, 1.2, 1.5, 1.6, 2.0\}\ \mathrm{g\,m^{-3}}$ \\[4pt]
    Initial mean droplet radius
      & $\bar{r}_0$ 
      & $\{9, 10, 11, 12, 13, 14, 15\}\ \mathrm{\mu m}$ \\[2pt]
    Gamma shape parameter
      & $\nu$ 
      & $\{0, 0.5, 1, 1.5, 2, 2.5, 3, 3.5, 4\}$ (unitless) \\
    \bottomrule
  \end{tabular}
\end{table}

\newpage

\section{Traditional Bulk parameterizations}\label{appendix:bulk}
Collision--coalescence leads to the generation of rain in clouds and is typically represented using a simplified bulk parameterization (Figure \ref{fig:overview}a). In a typical two-moment bulk representation for warm-rain cloud microphysics, the droplet size distribution in each ESM model grid box is divided into cloud (non-precipitating) and rain (precipitating) droplet populations. Each population is represented by its number and mass mixing ratios, giving four prognostic variables $(N_c,L_c,N_r,L_r)$ that describes the microphysical state of the cloud; here $N$ refers to number and $L$ refers to mass, with $c$ denoting cloud and $r$ denoting rain populations. In the two category, two moment representation, the collision--coalescence process is four separate process rates: autoconversion (AU), which transfers mass and number from the cloud to the rain category; accretion (AC), in which raindrops collect cloud droplets; and self-collection within the cloud (SC$_{c}$) and rain (SC$_{r}$) categories. Bulk parameterizations express these four process rates as a set of coupled ODEs that are functions of the four prognostic moments and any prescribed properties of the droplet size distribution, while conserving total mass. Single moment schemes (such as Kessler, 1969) use only the two mass moments as representations of the cloud and rain populations.

In Table \ref{tab:schemes}, we provide a summary of the 4 traditional bulk parameterizations for collision-coalescence that we re-implemented as differentiable code in JAX, and that are used as a comparison for data-driven and agentic AI parameterization development approaches. This list of bulk parameterizations for collision-coalescence is by no means comprehensive, but was selected to provide a range of parameterizations of varying sophistication and complexity. Kessler, 1969 is the simplest and predicts only the mass mixing ratios of the cloud and rain populations; Seifert and Beheng, 2001 is the most complex. All 4 parameterizations are still regularly used in atmospheric models.

\begin{table*}[ht]
  \centering
  \small
  \setlength{\tabcolsep}{6pt}
  \renewcommand{\arraystretch}{1.15}

  \caption{Implemented bulk collision--coalescence schemes.
  AU denotes autoconversion, AC accretion, and SC$_r$ and SC$_c$
  rain and cloud self-collection, respectively. Moment order denotes
  the number of predicted moments per hydrometeor category. The common
  box-model interface uses $(N_c,L_c,N_r,L_r)$; schemes that do not
  predict number moments use diagnostic or prescribed treatments.}
  \label{tab:schemes}

  \begin{tabular}{
    @{}
    >{\raggedright\arraybackslash}p{5.0cm}
    >{\centering\arraybackslash}p{2.5cm}
    >{\raggedright\arraybackslash}p{5.0cm}
    @{}
  }
    \toprule
    Reference & Moment order & Processes \\
    \midrule

    \cite[]{kessler1969distribution}
      & 1
      & AU, AC \\
    \addlinespace[3pt]

    \cite[]{berry1974analysis}
      & 2
      & AU, AC \\
    \addlinespace[3pt]

    \cite[]{khairoutdinovNewCloudPhysics2000}
      & 2
      & AU, AC \\
    \addlinespace[3pt]

    \cite[]{seifert2001double}
      & 2
      & AU, AC, SC$_r$, SC$_c$ \\

    \bottomrule
  \end{tabular}
\end{table*}

Below, we provide the original equations for the 4 reference parameterizations, and highlight in red the parameters which are retuned against the McSnow simulations (training data set).
\paragraph{Kessler (1969):} Threshold autoconversion, with $\mathrm{d}L_r/\mathrm{d}t=k_1(L_c-a)$
        for $L_c>a$, where $k_1=10^{-3}\,\mathrm{s^{-1}}$ and $a=0.5\,\mathrm{g\,m^{-3}}$, and accretion
        $k_2 E N_0^{1/8} L_c L_r^{7/8}$ with $k_2=6.96\times10^{-4}$, $E=1$ and $N_0=10^{7}\,\mathrm{m^{-4}}$
        \citep[Table~4]{kessler1969distribution}. Kessler works in water content rather than mixing
        ratio, so in SI units $k_{\text{ac}}=2.20$ and no air density appears. $N_c$ is treated diagnostically using a fixed mean mass $\bar{x}_c$, while $N_r$ is not predicted. This provides the simplest reference scheme. Four parameters are retuned against the McSnow training set.
\begin{align}
  \text{autoconversion:} \quad
  P_{\text{au}} &= \fit{k_{\text{au}}} \, \max\!\big(0,\; \Lc - \fit{L_{c0}}\big) \\
  \text{accretion:} \quad
  P_{\text{ac}} &= \fit{k_{\text{ac}}} \, \Lc \, \Lr^{\fit{\beta}} \\
  \frac{\diff \Lr}{\diff t} &= P_{\text{au}} + P_{\text{ac}},
  \qquad
  \frac{\diff \Lc}{\diff t} = -\frac{\diff \Lr}{\diff t},
  \qquad
  \frac{\diff \Nc}{\diff t} = \frac{\diff \Lc}{\diff t}\,\frac{\Nc}{\Lc},
  \qquad
  \frac{\diff \Nr}{\diff t} = 0 .
\end{align}
Kessler predicts no number concentrations; we add a proportional cloud-number sink, which holds the mean drop mass $\xbarc$ constant. The rain number concentration $\Nr$ is not predicted, so no number parameters appear.

\paragraph{Berry \& Reinhardt (1974), two-moment form:} \citet{berry1974analysis} derived autoconversion from numerical
        solutions of the collection equation; their rate equations are given in Part~IV of the series \citep{berry1974analysis4}.
        The scheme implemented here is not those rate equations but the corresponding two-moment form for the Long (1974) kernel,
        $\left.\mathrm{d}L_r/\mathrm{d}t\right|_{\mathrm{AU}}=C_{\mathrm{AU}}L_c^{2}\bar{x}_c^{2}$ with accretion $k_{\mathrm{cr}}L_cL_r$
        \citep[Eqs.~16 and 21]{seifert2001double}: the limit of Seifert \& Beheng without the universal functions that describe
        the feedback from existing rain on autoconversion and accretion. 
Three parameters are optimized:
\begin{align}
  P_{\text{au}} &= \fit{C_{\text{au}}} \; \Lc^2 \, \xbarc^{\,2} \\
  P_{\text{ac}} &= \fit{k_{\text{cr}}} \; \Lc \, \Lr \\
  \frac{\diff \Lr}{\diff t} &= P_{\text{au}} + P_{\text{ac}},
  \qquad
  \frac{\diff \Nr}{\diff t} = \frac{P_{\text{au}}}{\fit{x^*_r}},
  \qquad
  \frac{\diff \Nc}{\diff t} = -\frac{\nu_2}{\nu_1}\frac{P_{\text{au}}}{\xbarc}
                              - \frac{P_{\text{ac}}}{\xbarc},
\end{align}
with the shape-parameter ratio $\nu_2/\nu_1 = 3/2$ ($\nu_c = 1$) held fixed. $\fit{x^*_r}$ is the mass of a newly-formed raindrop.

\paragraph{Khairoutdinov \& Kogan (2000)} Empirical power-law fits to large-eddy simulations of marine stratocumulus, including
        $\mathrm{d}q_r/\mathrm{d}t =1350\,q_c^{2.47}(N_c/10^{6})^{-1.79}$
        for autoconversion and $67\,(q_cq_r)^{1.15}$ for accretion. New drizzle drops have radius $r_0=25\,\mu$m, and the drops lost from the cloud have the mean cloud-drop mass \citep[Eqs.~32 and 35]{khairoutdinovNewCloudPhysics2000}; the scheme has no rain self-collection. The scheme is based on regression within a specific cloud regime rather than an explicit process-rate theory. Six parameters are retuned: both prefactors, all three power-law exponents and the new-drop mass $x_{r0}$.
\begin{align}
  P_{\text{au}} &= \rho \; \fit{A_{\text{au}}} \; q_c^{\,\fit{a}} \; \big(10^{-6} \Nc\big)^{\fit{b}} \\
  P_{\text{ac}} &= \rho \; \fit{A_{\text{ac}}} \; \big(q_c \, q_r\big)^{\fit{c}} \\
  \frac{\diff \Lr}{\diff t} &= P_{\text{au}} + P_{\text{ac}},
  \qquad
  \frac{\diff \Nc}{\diff t} = -\frac{P_{\text{au}} + P_{\text{ac}}}{\xbarc},
  \qquad
  \frac{\diff \Nr}{\diff t} = \frac{P_{\text{au}}}{\fit{x_{r0}}} .
\end{align}

\paragraph{Seifert \& Beheng (2001)} The most complete scheme considered here. It combines autoconversion for the Long (1974) kernel with the universal function $\Phi_{\mathrm{au}}(\tau)$ describing the cloud--rain transition, accretion modified by $\Phi_{\mathrm{ac}}(\tau)$,
and cloud and rain self-collection. It predicts all four moments directly and can become stiff at high $L_c$.
Eight fitted parameters are retuned, including the two constants of the autoconversion universal function $\Phi_{\text{au}}$ and the
one in the accretion function $\Phi_{\text{ac}}$.
\begin{align}
  \Phi_{\text{au}}(\tau) &= \fit{\alpha} \, \tau^{\fit{\beta}} \big(1 - \tau^{\fit{\beta}}\big)^3 \\
  P_{\text{au}} &= \fit{C_{\text{au}}} \, \Lc^2 \, \xbarc^{\,2}
                   \left[ 1 + \frac{\Phi_{\text{au}}(\tau)}{(1-\tau)^2} \right] \\
  \Phi_{\text{ac}}(\tau) &= \left( \frac{\tau}{\tau + \fit{\tau_0}} \right)^{4} \\
  P_{\text{ac}} &= \fit{k_{\text{cr}}} \, \Lc \, \Lr \, \Phi_{\text{ac}}(\tau) \\
  \frac{\diff \Lr}{\diff t} &= P_{\text{au}} + P_{\text{ac}},
  \qquad
  \frac{\diff \Lc}{\diff t} = -\frac{\diff \Lr}{\diff t} \\
  \frac{\diff \Nc}{\diff t} &= \underbrace{-\,\fit{C_{\text{sc}}} \, \Lc \, \Nc \, \xbarc}_{\text{autoconv.\ + cloud self-collection}}
                              \underbrace{-\frac{P_{\text{ac}}}{\xbarc}}_{\text{accretion}} \\
  \frac{\diff \Nr}{\diff t} &= \frac{P_{\text{au}}}{\fit{x^*}}
                              - \underbrace{\fit{k_{\text{rr}}} \, \Nr \, \Lr}_{\text{rain self-collection}}
\end{align}
Following \citet[Appendix~A]{seifert2001double}, autoconversion removes $2P_{\text{au}}/x^*$ cloud droplets and cloud self-collection is defined so that the two processes together remove $k_c\frac{\nu+2}{\nu+1}L_c^2$, with $C_{\text{sc}}=k_c\frac{\nu+2}{\nu+1}$. Accretion and rain self-collection use the same Long-kernel constant, $k_r=5.78\,\mathrm{m^3\,kg^{-1}\,s^{-1}}$; here $k_{\text{cr}}$ and $k_{\text{rr}}$ are retuned separately.

\section{Retuned parameters in traditional bulk schemes against the McSnow training data set}

Table \ref{tab:fitparameters} provides the retuned parameters in the 4 bulk parameterizations after the optimization against the McSnow training data set. Table \ref{tab:fitloss} gives the training loss and held-out test loss for the 4 bulk parameterizations using their original published parameters and their retuned parameters, scored against the McSnow training and test data sets, respectively.

\begin{table*}[h]
\centering
\caption{Parameters for the 4 bulk reference schemes, retuned after optimizing against the McSnow training data set.}
\small
\begin{tabular}{llrrr}
\toprule
Scheme & Parameter & Published & Fitted & Ratio \\
\midrule
Kessler
 & $k_{\text{au}}$   & $1.00\times10^{-3}$ & $6.48\times10^{-5}$ & $0.065$ \\
 & $L_{c0}$          & $5.00\times10^{-4}$ & $9.09\times10^{-4}$ & $1.82$ \\
 & $k_{\text{ac}}$   & $2.201$             & $2.063$             & $0.94$ \\
 & $\beta$           & $0.875$             & $0.979$             & $1.12$ \\
\midrule
Berry \& Reinhardt
 & $C_{\text{au}}$   & $6.81\times10^{18}$ & $6.80\times10^{18}$ & $1.00$ \\
 & $k_{\text{cr}}$   & $5.78$              & $2.848$             & $0.49$ \\
 & $x^*_r$           & $2.60\times10^{-10}$& $3.53\times10^{-9}$ & $13.6$ \\
\midrule
Khairoutdinov \& Kogan
 & $A_{\text{au}}$   & $1350$              & $2919$              & $2.16$ \\
 & $a$               & $2.47$              & $2.694$             & --- \\
 & $b$               & $-1.79$             & $-3.219$            & --- \\
 & $A_{\text{ac}}$   & $67.0$              & $801$               & $12.0$ \\
 & $c$               & $1.15$              & $1.206$             & --- \\
 & $x_{r0}$          & $6.54\times10^{-11}$& $2.52\times10^{-10}$& $3.85$ \\
\midrule
Seifert \& Beheng
 & $C_{\text{au}}$   & $6.81\times10^{18}$ & $1.21\times10^{18}$ & $0.18$ \\
 & $\alpha$          & $600$               & $25390$             & $42.3$ \\
 & $\beta$           & $0.68$              & $0.933$             & --- \\
 & $k_{\text{cr}}$   & $5.78$              & $3.814$             & $0.66$ \\
 & $\tau_0$          & $5.00\times10^{-4}$ & $4.88\times10^{-5}$ & $0.098$ \\
 & $k_{\text{rr}}$   & $5.78$              & $3.857$             & $0.67$ \\
 & $C_{\text{sc}}$   & $1.42\times10^{10}$ & $7.00\times10^{9}$  & $0.50$ \\
 & $x^*$             & $2.60\times10^{-10}$& $3.21\times10^{-9}$ & $12.4$ \\
\bottomrule
\label{tab:fitparameters}
  
\end{tabular}
\end{table*}

\begin{table}[h]
\caption{Training loss and test loss scored against the McSnow reference data sets for the published and retuned parameters for the 4 reference schemes.}
\centering
\small
\begin{tabular}{lrrrr}
\toprule
 & \multicolumn{2}{c}{Train loss} & \multicolumn{2}{c}{Held-out test loss} \\
\cmidrule(lr){2-3}\cmidrule(lr){4-5}
Scheme & published & fitted & published & fitted \\
\midrule
Kessler                 & 3.127 & 0.800 & 3.988 & 0.776 \\
Berry \& Reinhardt      & 1.737 & 0.733 & 2.301 & 0.739 \\
Khairoutdinov \& Kogan  & 4.556 & 0.645 & 6.021 & 0.632 \\
Seifert \& Beheng       & 1.096 & 0.283 & 1.464 & 0.272 \\
\bottomrule
\label{tab:fitloss}
\end{tabular}
\end{table}

\newpage
\clearpage

\section{Metrics used for evaluation of bulk parameterization performance}\label{section:metrics}
The paper uses three metrics to evaluate the different collision-coalescence parameterizations: relative $L_2$ error, MAPE, and $t_{10}$. 

\begin{itemize}[leftmargin=*]
\item \textbf{Relative $L_2$ error}:
for each test trajectory and moment, $\|\hat{x}-x\|_2 / \|x\|_2$ over the time
series. The reported value is the median over test trajectories. Rain-mass error in Section~3 is the relative $L_2$ error of $L_r$.
\item \textbf{MAPE} (Figure~2b): for each trajectory,
$100 \times \mathrm{mean}_t\, |\hat{x}_t - x_t| / |x_t|$, over time steps with
$|x_t| > 10^{-20}$ in SI units; reported as the median over the 100 test
trajectories. 
\item \textbf{$t_{10}$}: time at which 10\% of the initial cloud mass has been
converted to rain. We report the median absolute error over test
trajectories, in minutes.
\end{itemize}

\section{Expert-in-the-loop effort}
Table \ref{tab:interventions} provides information about the number of prompts required for each of the 4 expert-in-the-loop modeling tasks described in the main text, as well as the approximate amount of time that each model component required for active development. These were determined from Claude's history (\texttt{history.jsonl}). 
\begin{table}[ht!!]
  \centering
  \small
  \caption{\textbf{Expert-in-the-loop effort for each agent-developed component},
  counted from the prompt history and corroborated by the git repo commit history. Substantive refer to prompts that direct work (excluding bookkeeping such as git commits), while corrective refers to prompts pointing out defects or mismatches. Active refers to total estimated time spent actively working on that component.}
  \label{tab:interventions}
  \begin{tabular}{>{\raggedright\arraybackslash}p{4.2cm}cccccc}
    \toprule
    Component & Sessions & Prompts & Substantive & Corrective
      & Active (h) & Commits \\
    \midrule
    Translate 4 literature schemes   &  1 &   5 &   4 &  2 &  0.5 &  3 \\
    Seifert \& Rasp (2020) MLPs      &  3 &  25 &  19 &  4 &  4.0 &  2 \\
    SuperdropNet (2025)              &  3 &  36 &  24 &  8 &  7.1 &  3 \\
    Physics-constrained Neural ODE   &  5 & 131 & 101 & 14 & 22.0 & 19 \\
    \bottomrule
  \end{tabular}
\end{table}

\section{Physics-constrained Neural ODE closure}\label{app:node}

To explore whether agentic AI could come up with genuinely new data-driven parameterization approaches for collision-coalescence,  we explored expert-in-the-loop development of a Neural ODE-based \cite[]{chenNeuralOrdinaryDifferential2019} closure for the evolution of the four two-moment state variables. Rather than proposing a generic black-box network, Claude drew on structural ideas from the historical bulk schemes to factor the learned tendencies into physically meaningful process rates. The architecture enforced mass conservation by construction and was designed so that its learned components could be associated with recognizable transfers between cloud and rain categories. These choices produced a more interpretable and physically constrained model than an unconstrained state-to-state neural network.

Developing the closure nevertheless required substantial expert intervention (Table \ref{tab:interventions}), with 131 prompts over 22 active hours. Multiple iterations were needed to refine the approach, with design iterations focused on architecture refinement and training objective exploration dominating the exploration phase. Although the resulting NODE captured the qualitative evolution of the McSnow trajectories, it remained less accurate than several of the reconstructed classical and machine-learning parameterizations (Section \ref{sec:evaluation}). Given the extent of the expert intervention, we interpret this experiment as human--agent co-development rather than autonomous parameterization discovery. It shows that an agent can help explore architectures that combine learned components with known physical structure, but not that agentic development will necessarily produce a more accurate closure.

\section{Reconstruction of machine-learning baselines}
\label{app:ml_reconstruction}

For both reconstructions, the coding agent was given the original publication and the McSnow training dataset and was guided through implementation, training, and comparison with the published results. Table \ref{tab:interventions} summarizes the expert involvement required for each reconstruction.

\begin{figure}[ht]
\noindent\includegraphics[width=\textwidth]{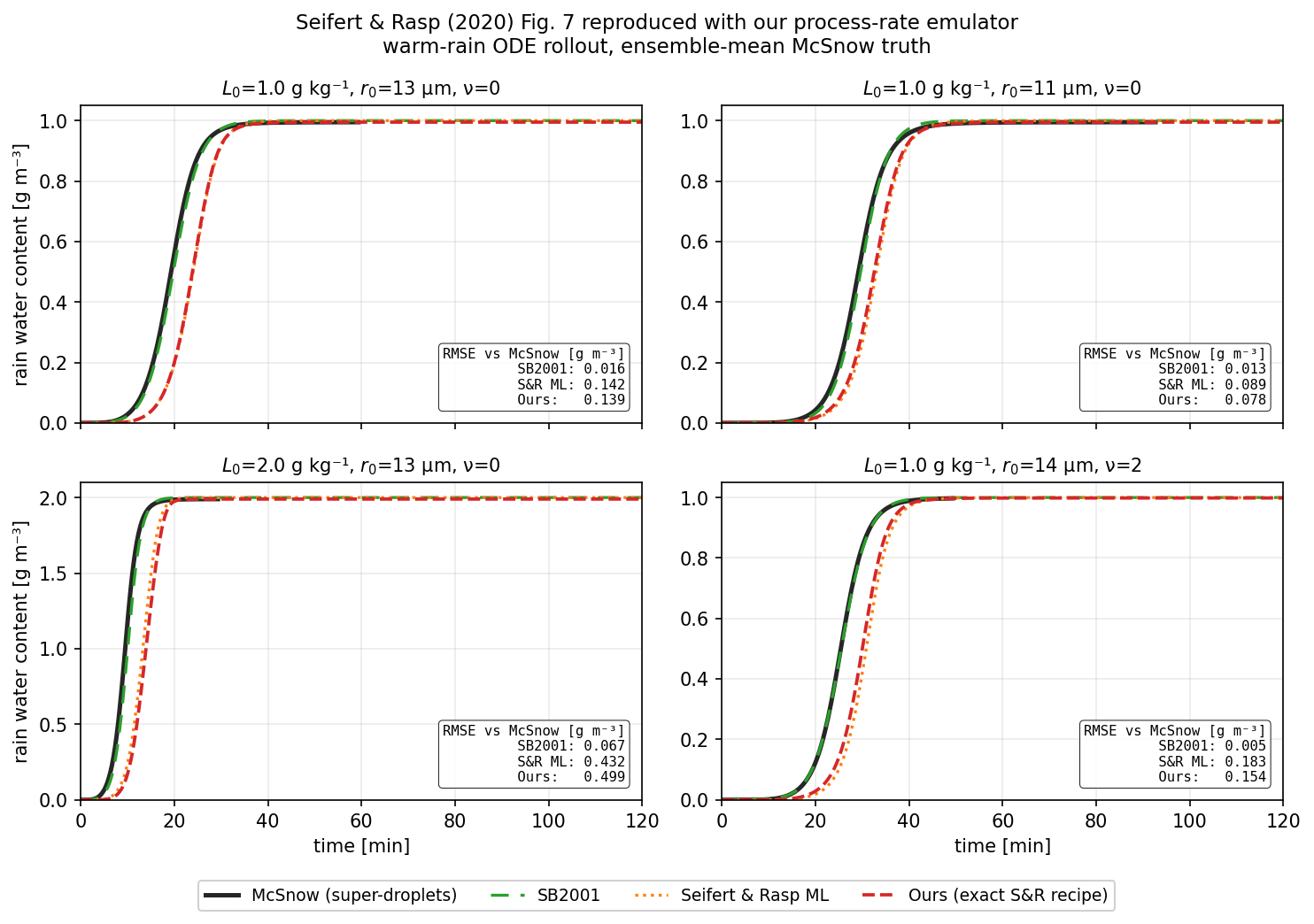}
\caption{\textbf{Reconstruction of Figure 7 from Seifert and Rasp, 2020.}}
\label{fig:seifertFig7}
\end{figure}

\paragraph{Seifert and Rasp (2020).}
The parameterization developed by \citet{seifert2020potential} uses four separate multilayer perceptrons to predict rates of autoconversion, accretion, cloud self-collection, and rain self-collection. These rates define the right-hand side of the ODEs governing the four bulk moments, which are integrated using a forward Euler scheme. The agent implemented the networks and training procedure in JAX and reproduced the published evaluation framework. The reconstructed model closely reproduced both the single-step rate predictions and prognostic behavior of the original model. When initialized from the same conditions, its rollouts also reproduced the reported behavior relative to the McSnow simulations and the Seifert and Beheng parameterization (Figure \ref{fig:seifertFig7}). The reconstruction required three development sessions, 19 substantive prompts, four corrective prompts, and approximately 4.0 hours of active expert involvement (Table \ref{tab:interventions}).

\paragraph{SuperdropNet.}
SuperdropNet \citep[]{sharma_superdropnet_2024} predicts tendencies for the four bulk moments rather than separate conventional process rates and applies physical constraints to the resulting state updates. Its training procedure progressively increases the autoregressive rollout horizon from one to 25 time steps, with the model from each stage used to initialize the next.

\begin{figure}[ht]
\noindent\includegraphics[width=\textwidth]{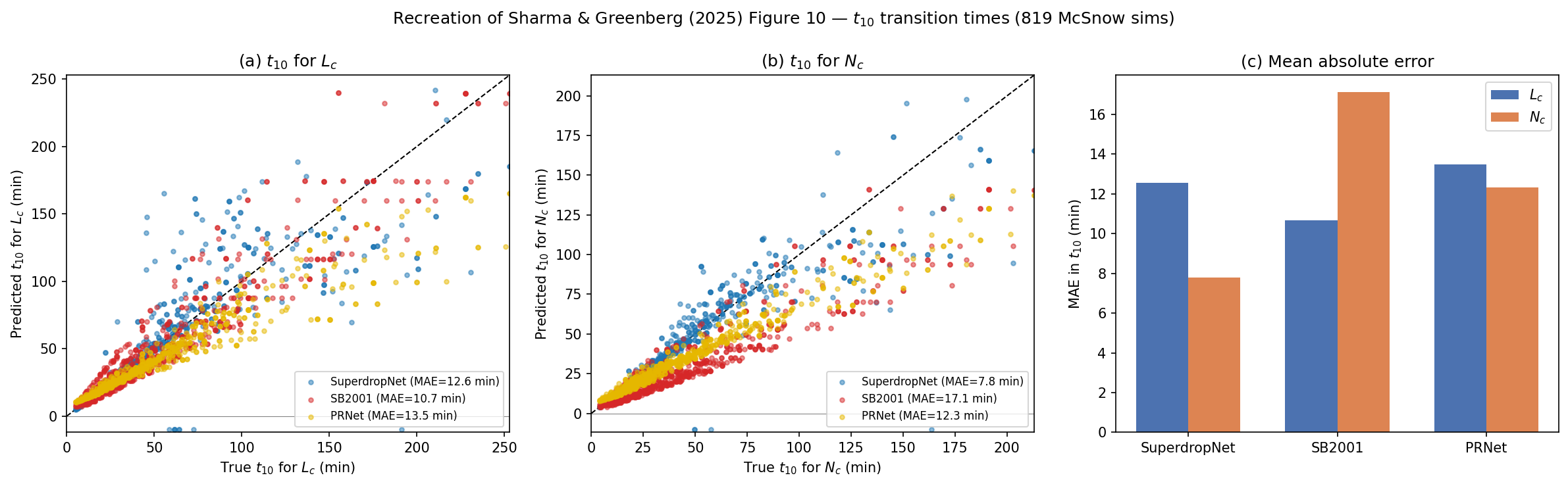}
\caption{\textbf{Reconstruction of Figure 10 from Sharma and Greenberg, 2025.}}
\label{fig:superdropnetFig10}
\end{figure}

The agent reconstructed the network architecture and overall loss function, but reproducing the published training performance proved significantly more difficult. The resulting model captured the qualitative behavior of SuperdropNet but did not achieve the same accuracy during autoregressive rollout (Figure \ref{fig:superdropnetFig10}). The main difficulty was reproducing the complete optimization procedure, including the learning-rate schedule, curriculum transitions, and stopping criteria. The reconstruction required three development sessions, 24 substantive prompts, 8 corrective prompts, and approximately 7.1 hours of active expert involvement (Table \ref{tab:interventions}). These examples show that agentic AI can reliably reconstruct relatively simple ML parameterizations when the architecture and training objective are clearly specified. Reproducing more complex models with autoregressive or curriculum-based training remains considerably more dependent on expert intervention.

\section{Single shot LLM calls and iterative refinement experiments}\label{appendix:models}

\subsection{Models used in experiments}
Claude Opus~4.8, used for the expert-guided experiments, was not available through the American Science Cloud model gateway that was used for the majority of the single shot and iterative refinement experiments. Claude Opus~5.5, a later model in the same family, is used as the closest same-family comparison for the unattended experiments.

\paragraph{Cost of the unattended experiments.} Here we estimate token costs for the single-shot and iterative experiments, priced at the provider's public standard-tier rate. 

\begin{table}[h]
\centering\small
\begin{tabular}{lrrrrr}
\toprule
Model & Translation & Single-shot & Iterative & Iterative & Total \\
 & (40 calls) & (10 calls) & (10$\times$4 rounds) & (5$\times$12 rounds) & \\
\midrule
Gemini 2.5 Pro & 5.62$^{\dagger}$ & 1.77$^{\dagger}$ & 4.15$^{\dagger}$ & -- & 11.54 \\
Gemini 3.1 Pro (preview) & 5.48$^{\ddagger}$ & 3.65 & 13.23 & -- & 22.37 \\
Gemini 4 Argon & 1.95$^{\S}$ & 0.28$^{\S}$ & 2.22$^{\S}$ & 1.75$^{\S}$ & 6.21 \\
Claude Haiku 5.5 & 0.66 & 0.08 & 0.15 & -- & 0.89 \\
Claude Sonnet 5.5 & 3.72 & 3.48 & 9.15 & -- & 16.34 \\
Claude Opus 5.5 & 6.14 & 5.55 & 10.16 & 6.50 & 28.34 \\
GPT-5.6 Terra & 2.60 & 0.83 & 1.90 & -- & 5.33 \\
GPT-6 Luna & 0.14 & 0.04 & 0.10 & -- & 0.29 \\
GPT-6 Sol & 1.84 & 1.03 & 2.01 & -- & 4.88 \\
GPT-6 Astra & 8.11 & 4.86 & 7.96 & -- & 20.92 \\
\midrule
Total & 36.27 & 21.57 & 51.03 & 8.24 & 117.11 \\
\bottomrule
\end{tabular}
\caption{Cost (USD) of the unattended experiments per model, at each provider's
public standard-tier API prices (October 2026). Thinking tokens are billed as output
and cached input at the cache-read rate. Each cell is the mean cost per call times the number of calls in the column header (one set of experiments, without retries).
Runs were made through the AmSC Model Access Gateway, so these are list-price
equivalents. $^{\dagger}$Usage was not recorded for the original Gemini~2.5 Pro runs
(called through the Gemini API); estimated from a re-measured sample of the same
calls (4 translations, 2 single-shot, 1 iterative candidate). $^{\ddagger}$The 10 Berry \& Reinhardt translations were made through the Gemini API
and are estimated from the mean Gemini~3.1 Pro translation call.
$^{\S}$Gemini~4 Argon was run at its default setting, without reasoning effort
set, and used almost no thinking tokens; with thinking enabled its cost would be
substantially higher.}
\label{table:models}
\end{table}

\subsection{Summary of experiments with different models}
Here we provide results for the single shot API calls and automated LLM closure generation with iterative feedback. We compare performance across 10 different large language models. Repeating the unattended experiments across models separates the effect of model choice from that of the workflow, and reveals failure modes common to all models.

\begin{figure}[ht]
\noindent\includegraphics[width=1.0\textwidth]{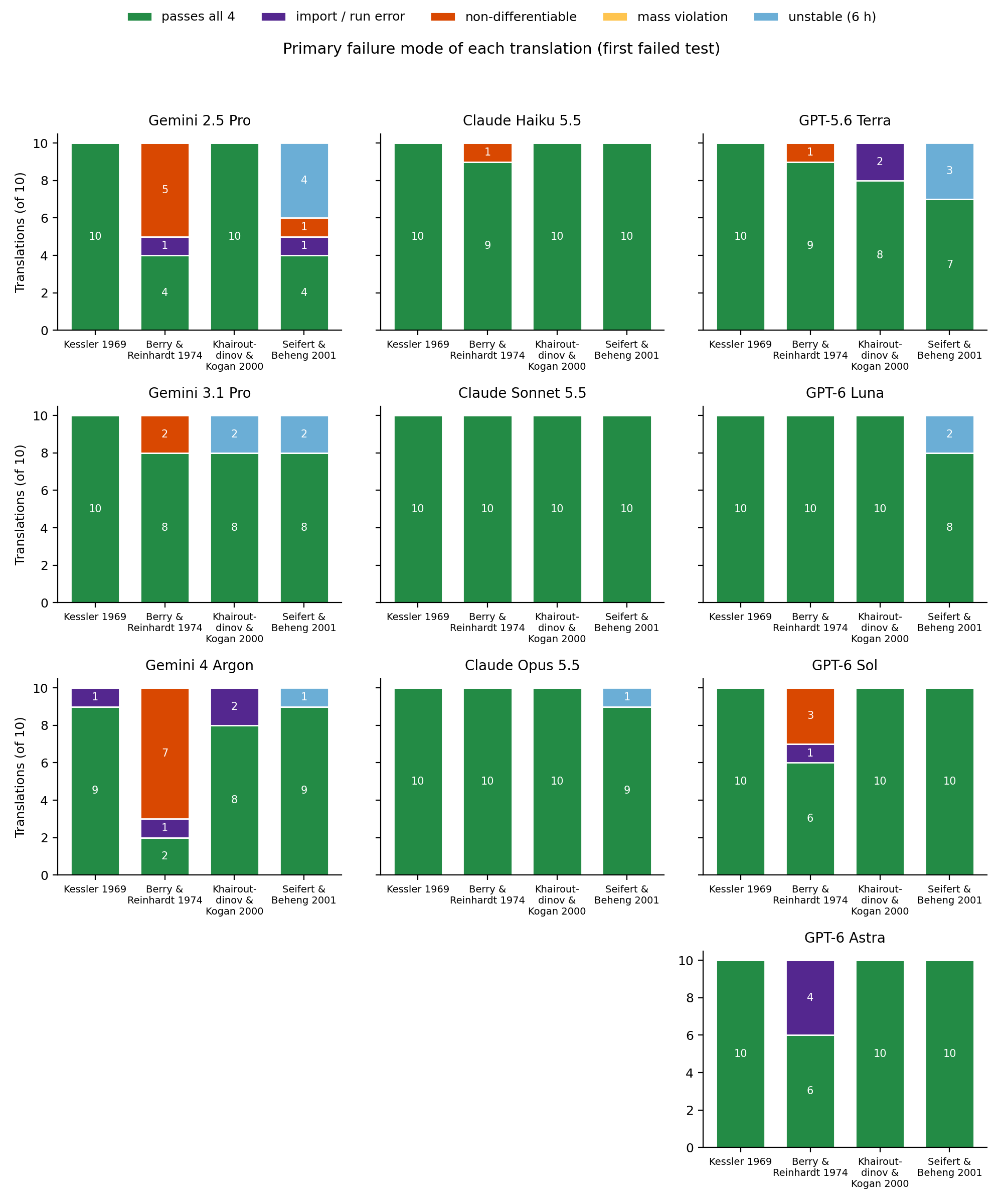}
\caption{\textbf{Failure modes for single-shot translation of historical schemes by model.}}
\label{fig:sshistorical}
\end{figure}

\begin{figure}[ht]
\noindent\includegraphics[width=1.0\textwidth]{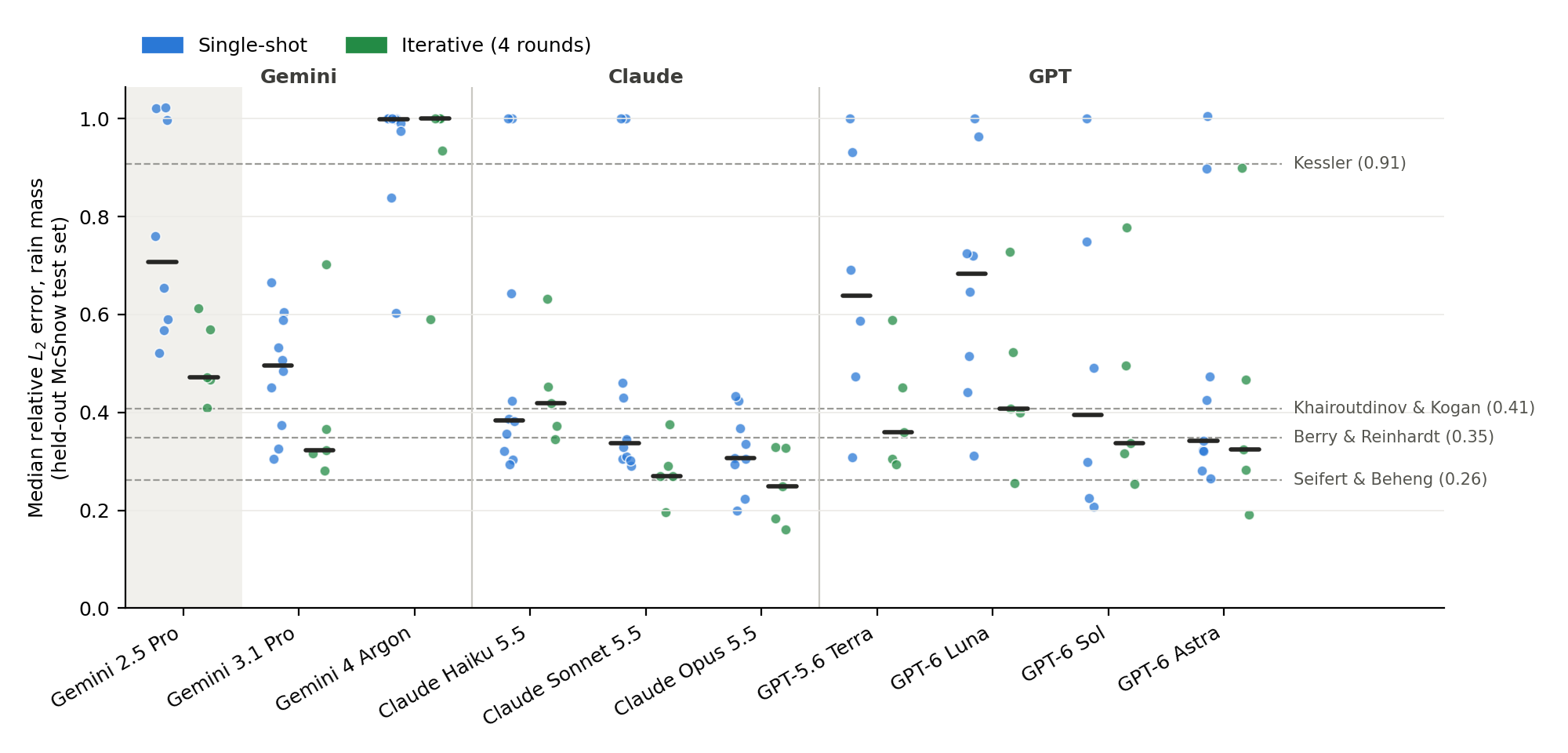}
\caption{\textbf{Accuracy of single-shot and iteratively refined agent-generated closures relative to the McSnow test data and historical parameterizations (shown as dashed lines scored against the McSnow test data).} Here, the performance for the historical parameterizations assumes their published (rather than retuned) parameters.}
\label{fig:iterativecomparison}
\end{figure}

\begin{figure}[ht]
\noindent\includegraphics[width=1.0\textwidth]{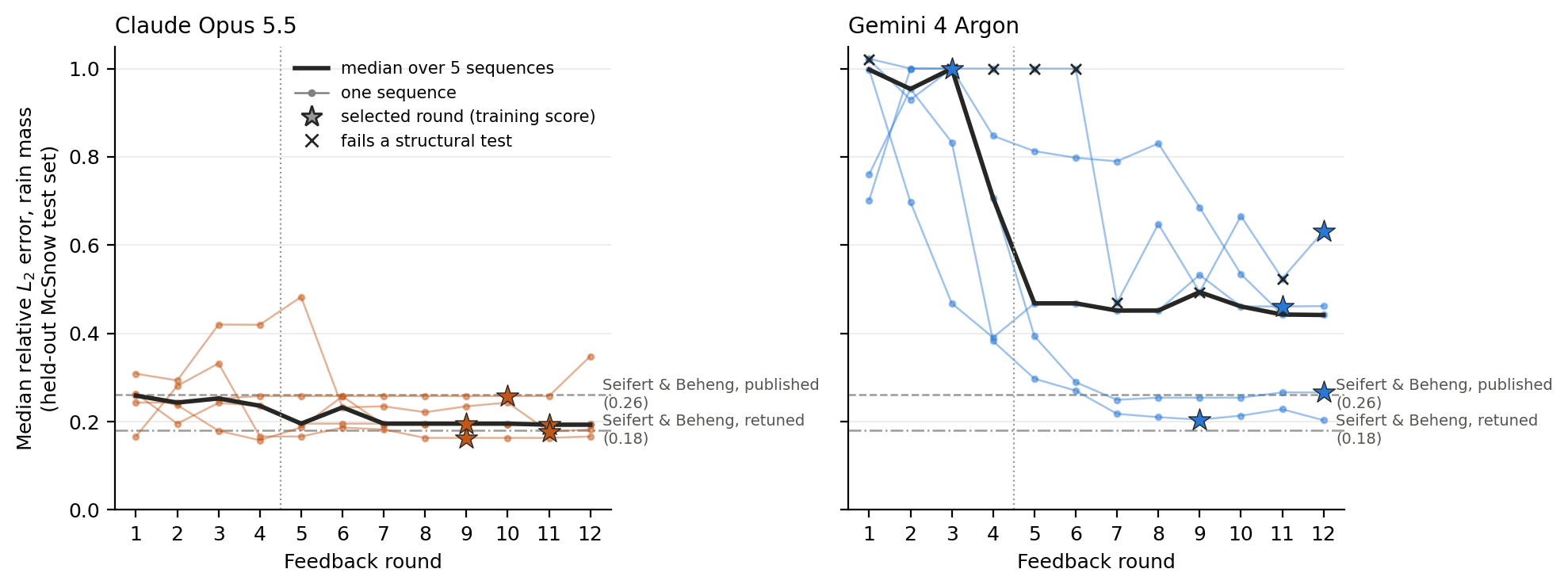}
\caption{\textbf{Accuracy of iteratively refined agent-generated closures relative to the McSnow test data and historical parameterizations for the 12 round iterative refinement rounds.}}
\label{fig:iterativecomparison12rounds}
\end{figure}

\clearpage
\newpage

\paragraph{Best agent-generated closure vs Seifert \& Beheng.}
The best closure (Claude Opus~5.5, iterative sequence~01) is a modified Seifert \& Beheng scheme rather than a new one. A comparison between the schemes is shown in Table \ref{tab:opus_vs_sb}.
\begin{table}[h]
\centering\footnotesize
\begin{tabularx}{\textwidth}{@{}>{\raggedright\arraybackslash}p{2.5cm} X X >{\raggedright\arraybackslash}p{1.5cm}@{}}
\toprule
Process & Seifert \& Beheng (2001) & Claude Opus~5.5 closure & Relation \\
\midrule
Autoconversion, moment dependence & $C_{au} L_c^2 x_c^2$ & $K_{au} L_c^2 x_c^2$ & Same form \\
Autoconversion, rain-fraction correction & $1 + \Phi_{au}(\tau)/(1-\tau)^2$, \newline
$\Phi_{au} = 600\,\tau^{0.68}(1-\tau^{0.68})^3$ &
$\epsilon + \tau/[1 + (\tau/\tau_s)^{0.6}]$, \newline $\tau_s = 7.3\times10^{-5}$ &
Different \\
Accretion, kernel & $K_{cr} L_c L_r$, $K_{cr} = 5.78$ & $K_{ac} L_c L_r$, $K_{ac} = 5.5$ &
Same form \\
Accretion, correction & $[\tau/(\tau + 5\times10^{-4})]^4$ (rain fraction) &
$(x_c/1.3\times10^{-11}\,\mathrm{kg})^{0.25}$ (cloud-drop mass) & Different \\
Cloud number, autoconversion + self-collection & $\tfrac{\nu+2}{\nu+1}K_{cc} L_c^2$, $K_{cc} = 9.44\times10^9$, $\nu=1$ & $K_{cc} L_c^2 + P_{au}/x_c$ & Different split \\
Rain self-collection & $-K_{rr} N_r L_r$, $K_{rr} = 5.78$ &
$-5.78\,N_r L_r \big/$ $[1 + (x_r/2.6\times10^{-6}\,\mathrm{kg})^{0.68}]$ &
Same form, large drops limited \\
Mass of a new rain drop & $x^* = 2.6\times10^{-10}$\,kg & $5\times10^{-10}$\,kg &
Same role \\
Cloud number lost by accretion & $C/x_c$ & $0.8\,C/x_c$ & Rescaled \\
\midrule
$L_r$ relative $L_2$ error & 0.26 published, 0.18 retuned & 0.16 & \\
$t_{10}$ error [min] & 7.7 published, 6.0 retuned & 5.5 & \\
$L_r$ rollout MAPE [\%] & 191 published, 47 retuned & 48 & \\
FLOPs per step & 66 & 67 & \\
\bottomrule
\end{tabularx}
\caption{Process formulations of Seifert \& Beheng (2001) and of the most accurate
agent-generated closure (Claude Opus~5.5, iterative), and their errors on the 100
held-out McSnow test simulations. $\tau = L_r/(L_c+L_r)$ is the rain fraction and
$x_c$, $x_r$ the mean cloud- and rain-drop masses.}
\label{tab:opus_vs_sb}
\end{table}

\end{document}